\documentstyle[12pt]{article}
\begin{document}
\baselineskip 1.3em
\newcommand\ba{\begin{array}}
\newcommand\ea{\end{array}}
\newcommand\bc{\begin{center}}
\newcommand\ec{\end{center}}
\newcommand\be{\begin{enumerate}}  
\newcommand\ee{\end{enumerate}}  
\newcommand\bi{\begin{itemize}}  
\newcommand\ei{\end{itemize}}  
\newcommand\bd{\begin{description}}  
\newcommand\ed{\end{description}}  
\newcommand\beq{\begin{equation}}  
\newcommand\eeq{\end{equation}}  
\newcommand\beqa{\begin{eqnarray*}}  
\newcommand\eeqa{\end{eqnarray*}}  

\newcommand\gap{\smallskip}
\newcommand\eqbox[1]{\fbox{\rule[-.8em]{0em}{2.3em}$\displaystyle\ #1\ $}}

\newcommand{\dum}{\thechapter.}
\newcommand{\eq}[1]{Eq.\ (\ref{#1})}
\newcommand{\eqs}[2]{(\ref{#1}--\ref{#2})}
\newcommand\cf{{\em cf\ }}
\newcommand\eg{{\em e.g.,\ }}
\newcommand\etc{{\em etc}}
\newcommand\ie{{\em i.e.,\ }}
\newcommand\qed{\hfill {\bbox{ QED}}}

% The general mathematics macros
\newcommand\A{\op A}
\newcommand\B{\op B}
\newcommand\mathC{\mkern1mu\raise2.2pt\hbox{$\scriptscriptstyle|$}
                {\mkern-7mu\rm C}}
\newcommand\D{\op D}
\renewcommand\H{{\cal H}}                       %% Redefinition
\renewcommand\P{\op P}                          %% Redefinition
\newcommand\Q{\op Q}
\newcommand\U{\op U}
\newcommand\Z{{\rm \bf Z}}

\newcommand\abs[1]{\vert {#1}\vert}
\newcommand\av[1]{\langle#1\rangle}
\newcommand\bra[1]{\langle#1\vert\,}
\newcommand\bracket[2]{\langle#1\vert#2\rangle}
\newcommand\dby[1]{{d\over d#1}}
\def\ddde{\de^{(3)} }
\newcommand\down{\vert\!\downarrow\,\rangle}

\newcommand\half{{\frac 12}}
\newcommand\hhalf{{\textstyle\half}}
\newcommand\ioh{{i\over\hbar}}

\newcommand\ket[1]{\,\vert#1\rangle}
\newcommand\la{\langle}
\newcommand\lleft{\vert\!\leftarrow\rangle}
\newcommand\map{\longrightarrow}
\newcommand\mathR{{\rm I\! R}}
\newcommand\norm[1]{\parallel\!\v#1\!\parallel}
\newcommand\op[1]{\widehat{#1}}
\newcommand\ra{\rangle}
\newcommand\rright{\vert\!\rightarrow\rangle}
\newcommand\tr{{\rm tr}\,}
\newcommand\triple[3]{\langle#1\vert\,#2\,\vert#3\rangle}
\newcommand\twid[1]{\tilde {#1}}
\newcommand\unit{{\rm I}}
\newcommand\up{\vert\!\uparrow\,\rangle}
\renewcommand\v[1]{{\bf{#1}}}        %% Redefinition!

\newcommand\<[2]{\langle\v#1,\v#2}              
\renewcommand\>{\rangle}                        %% Redefinition!
\renewcommand\[{[\,}                            %% Redefinition!
\renewcommand\]{\,]}                            %% Redefinition!

% The greek letters
\renewcommand\a{\alpha}                         %% Redefinition!
\renewcommand\b{\beta}                          %% Redefinition!
\newcommand\g{\gamma}
\newcommand\de{\delta}
\newcommand\e{\varepsilon}
\newcommand\z{\zeta}
\newcommand\k{\kappa}
\renewcommand\l{\lambda}                        %% Redefinition!
\newcommand\m{\mu}
\newcommand\n{\nu}
\newcommand\r{\rho}
\newcommand\s{\sigma}
\newcommand\th{\theta}
\newcommand\f{\phi}
\newcommand\w{\omega}
\renewcommand\O{\Omega}				%% Redefinition!
\renewcommand\L{\Lambda}			%% Redefinition!
\newcommand\bolde{\epsilon\mkern-6mu\epsilon}  % Bold faced epsilon
\newcommand\nnabla{\nabla\mkern-14mu\nabla}  %  "     "   nabla
\newcommand{\mue}{{\mbox{\boldmath $\mu$}}}

\def\pdby#1{\partial\over\partial #1}
\def\exl{\raise1pt\hbox{$\scriptstyle<$}}
\def\exr{\raise1pt\hbox{$\,\scriptstyle>$}}

\newcommand\rem[1]{$\vert \underline{\overline{\text{{\bf#1}}}}\vert
$}
\newcommand\dmu{\partial_\mu}
\title{Derivative expansions of real-time thermal effective actions}
\author{Maria Asprouli, Victor Galan-Gonzalez\\
        \\
        Imperial College\\
	Theoretical Physics Group\\
	Blackett Laboratory\\
	London SW7 2BZ}
\maketitle

\begin{abstract}

In this work we use a generalised real-time path formalism with properly 
regularised propagators based on Le Bellac and Mabilat \cite{belmab}
and calculate the effective potential and the higher 
order derivative terms of the effective action in the case of real scalar 
fields at finite temperature. We consider time-dependent fields in thermal 
equilibrium and concentrate on the quadratic part of the expanded effective
 action which has been associated with problems of non-analyticity at the zero
limits of the four external momenta at finite temperature. We derive the 
effective potential and we explicitly show its independence of the initial 
time of the system when we include both paths of our time contour. We also 
derive the second derivative in the field term and recover the Real Time (RTF)
and the Imaginary Time Formalism (ITF) and show that the divergences 
associated with the former are cancelled as long as we set the regulators 
zero in the end. Using an alternative method we write the field in its Taylor 
series form and we first derive RTF and ITF in the appropriate limits, check
the analyticity properties in each case and do the actual time derivative 
expansion of the field up to second order in the end. We agree with our 
previous results and discuss an interesting term which arises in this 
expansion. Finally we discuss the initial time-dependence of the quadratic 
part of the effective action before the expansion of the field as well as 
of the individual terms after the expansion.

\end{abstract}

\section{Introduction}

The interest in the amalgamation of field theory and 
statistical mechanics arose from the realisation that many problems 
encountered experimentally and theoretically in particle physics have 
many-body aspects. For this reason, zero-temperature quantum field 
theory was reformulated by generalising the usual time-ordered products 
of operators to the ordered products along a path in a complex 
time-plane \cite{bel}. The choice of the path gives rise to different 
formalisms but all theories should give the same physical answers. 
Although the various path-ordered 
finite temperature field theory formalisms such as Real Time Formalism (RTF) 
including the closed-time approach \cite{Keld} and  
Imaginary Time Formalism (ITF) \cite{Matsu} should give the same physics, 
there has been serious discussion about their exact equivalence. 

In this paper we will tackle a problem in RTF which consists in the 
occurrence of pathologies associated with singularities, arising in 
diagrams with self-energy 
insertions or in some effective potential calculations. This problem appears
when products of delta functions with the same argument are involved and 
creates non-analyticities in the effective action at finite temperature, thus 
making it ill-defined.

The main interest for developing an effective formalism which describes 
the  finite temperature field theory comes from the need to tackle important 
problems in phase transitions, which have played a crucial role in the early 
evolution of the 
universe. The significance and observable quantities of a specific 
transition depend on its detailed nature and its order. A reason for a 
well-defined effective action comes from the fact that it
represents the quantum 
corrections which in general might be of extreme importance in defining or 
changing the order of a transition. For example, analytic analysis 
\cite{ander} suggests that the electroweak phase transition is first order 
because 
of quantum corrections from gauge bosons while non-perturbative lattice
simulations of high temperature electroweak theory suggests that this
is only true if the Higgs is lighter than 70Gev \cite{kaj}. The inclusion 
of higher derivative terms in the derivative expansion 
of the effective action in a first order transition is of great importance in 
cases such as the derivation
of the rate of the sphaleron fluctuations. These configurations have been 
used to explain the observed baryon asymmetry of the universe \cite{arno}.

Moreover, although the effective potential can give the approximate 
critical temperature of a given transition, it is not adequate for answering 
questions concerning the departure of the field from equilibrium occurring 
during dynamical cooling near and below the critical temperature $T_{c}$. The 
effective potential describes static properties and 
therefore it is not an appropriate tool for studying the dynamical 
behaviour of a wide class of field theory models considered in 
inflationary scenarios.

The standard method for estimating the quantum corrections is to first 
integrate out quantum fluctuations about a {\it constant} background. This
 gives 
an {\it effective potential} for $\phi$ which is then used in the equations of 
motion determining $\bar{\phi}(\bar{x},\tau)$ \cite{cole2}. 
Integrating out fluctuations about a general {\it inhomogeneous} 
configuration gives the full effective action which includes the higher 
derivative 
terms. In the language 
of quantum field theory at finite temperature the effective potential
${\it{\Gamma}}$ is given by ${\it{\Gamma}}=\beta F$, where $F$ is the minimum 
of the 
free energy at which the system lies in the case of local thermal 
equilibrium. If the ensemble averages of the matter fields are 
homogeneous and static, then the free energy is given by the finite 
temperature effective potential \cite{jac}. 

In analogy to quantum mechanics, the decay rate of an unstable configuration 
$\phi_{f}$ with energy ${\cal E}_f$ is given by \cite{cole1}

$$\Gamma=-\frac{2}{\hbar}\mbox{Im}\[{\cal E}_f\]=
-2\mbox{Im}\[\lim_{T \rightarrow \infty} 
\frac{1}{T}\ln \bra{\phi_f}e^{-HT/\hbar}\ket{\phi_f}\]
$$
where H is the Hamiltonian and the matrix element can be described as a 
functional integral

$$\bra{\phi_f}e^{-HT/\hbar}\ket{\phi_f}=
N\int{\cal D}\phi e^{-S[\phi]/ \hbar}
$$
in Euclidean time.
Here $\phi$ is subject to the condition $\phi(T/2)=\phi(-T/2)=\phi_{f}$ and 
$S$ denotes the Euclidean action. Evaluating the functional integral to 
one loop, we expand $S(\phi)$ about a solution of the equations of motion,
 $\bar\phi$, and keeping only terms quadratic in the fluctuations 
$\delta\phi=\phi-\bar\phi$, we obtain 

\beqa
N\int{\cal D}\phi e^{-S[\phi]/ \hbar} & \simeq & 
N e^{-S(\bar\phi/ \hbar)}\[\det(-\dmu\dmu+V''(\bar{\phi}))\]^{-\half}\\
&\equiv & \exp\[-S_{eff}(\bar{\phi}/\hbar)\]
\eeqa
where $S_{eff}$ is the effective action. If we expand $S_{eff}$ about a 
constant $\phi$, i.e. in powers of momentum about a point with zero external 
momenta, in position space and zero temperature this reads
$$
S_{eff}(\phi)=\int d^4x\[-V_{eff}(\phi)+
\half Z(\phi)\dmu\phi\partial^{\mu}\phi +{\cal O}((\dmu\phi)^4)\]
$$
where we have made use of ${\it{T}=0}$ Lorentz properties. For constant 
$\phi$ only the effective potential term survives.

Although such an expansion up to the second derivative has been performed 
at zero temperature for scalar and Dirac field theories \cite{cole3}, there 
are difficulties arising in the equivalent expansion at finite temperature. 
Das and Hott \cite{das} find a non-analyticity in the two-point functions 
involving 
the temperature dependent term of the quadratic part of the effective action. 
In this spirit, if the derivative expansion breaks down at finite temperature,
 the definition of the effective potential might not be unique. This 
non-analyticity, in the case of the vacuum polarisation for a scalar field 
coupled to a classical external field, manifests itself in the difference 
between the order of the limits $(p_{0}\rightarrow 0, 
{\bf p}\rightarrow 0)$ and 
$({\bf p}\rightarrow 0, p_{0}\rightarrow 0)$ of the external momenta, 
the first relating to the electric screening mass of the photon and 
the second to the plasma frequency of the particular field theory under 
consideration \cite{grib}. This non-commutativity appears in hot 
QCD \cite{sili}, self interacting scalars \cite{weld1,evans} and gauge 
theories with 
chiral fermions \cite{weld2}. In ITF, setting $p_{\mu}=0$ 
first and performing the mode sum gives the same result as taking the limit 
$p_0\rightarrow 0$  first and the limit ${\bf p}\rightarrow 0$ afterwards. 
In RTF extra Feynman rules 
have been imposed to explain this 
difference in the two limits \cite{eva1}. The problem is neither due to 
subtleties 
in the use of Feynman parametrisation at finite temperature \cite{weld3}, 
nor to the infinite number of possible extensions of $p_{0}$ to the 
imaginary axis and its analytic continuation to the complex plane \cite{evans}.
The lack of analyticity and the infrared divergences occurring in the 
definition of the effective action at finite temperature show the need 
for an effective field theory formalism from which RTF and ITF rules 
can be derived easily. In the next chapter we will describe a method 
dealing with these problems. We will calculate the
two-propagator contribution (bubble diagram) to the second derivative term
 in the effective action using Le Bellac and Mabilat's generalised real-time 
path formulation with properly regularised propagators \cite{belmab}.

\section{The method}

In Le Bellac and Mabilat's approach \cite{belmab} they derive Feynman rules 
that take explicitly into
account the vertical part of the contour and recover the RTF 
in the case of diagrams with at least one finite external line and the
ITF in the case of vacuum fluctuations. They keep the regulators of the
propagators when they find problematic products of delta functions with
the same argument and show that they can use RTF when no such problems
arise. They claim that the contribution of the vertical part of the 
contour lies in the 
cancellation of the $t_{i}$ dependent terms of the horizontal part since 
the whole result should be  $t_{i}$ and $t_{f}$ independent due to the 
KMS condition of the propagators. We will now describe in detail this
method, which we will use throughout this paper.

\subsection{Outline of the method}

The specific approach uses the Mills \cite{mill} mixed representation of the 
propagators for a free scalar field, with $t$ defined in the generalised time 
path C, starting at $t_{i}$ and ending at $t_{i}-i \beta$ of Fig.1. 
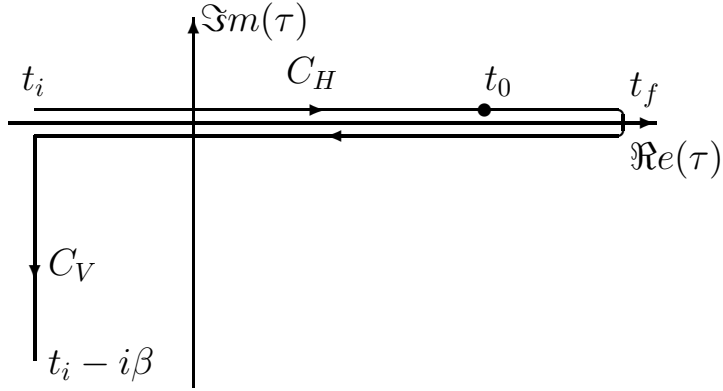
\begin{figure}[htb]
\begin{center}
\setlength{\unitlength}{0.5pt}
\begin{picture}(495,280)(35,480)
\put( 70,565){\makebox(0,0)[lb]{\large $C_V$}}
\put(185,750){\makebox(0,0)[lb]{\large $\Im m(\tau)$}}
\put(250,710){\makebox(0,0)[lb]{\large $C_H$}}
\put(510,645){\makebox(0,0)[lb]{\large $\Re e (\tau)$}}
\put(510,700){\makebox(0,0)[lb]{\large $t_f$}}
\put(400,705){\makebox(0,0)[lb]{\large $t_0$}}
\put( 50,705){\makebox(0,0)[lb]{\large $t_i$}}
\put(70,490){\makebox(0,0)[lb]{\large $t_i-i \beta$}}
\thicklines
\put(400,690){\circle*{10}}
\put( 40,680){\vector( 1, 0){490}}
 
\put( 60,690){\vector( 1, 0){220}}
\put(280,690){\line( 1, 0){220}}
\put(500,680){\oval(10,20)[r]}
\put(500,670){\vector(-1, 0){220}}
\put(280,670){\line(-1, 0){220}}
\put( 60,670){\vector( 0,-1){110}}
\put( 60,560){\line( 0,-1){ 60}}
\put(180,480){\vector( 0, 1){280}}
\end{picture}
\end{center}
\caption{The integration contour C in the complex t-plane.}
\end{figure}
The propagator is written as
\beq
\label{eqn1}
D_c(t,\v{k})=\int \frac{d k_0}{2\pi}e^{-i k_0
t}\[\th_c(t)+n(k_0)\]\rho(k_0,\v{k})
\eeq
where $\theta_{c}(t)$ is a contour $\theta$ function, $n(k_{0})$ is the 
Bose-Einstein distribution function given by
\beq
\label{eqn2}
n(k_0)=\frac1{e^{\b k_0}-1}
\eeq
 and $\rho(k_{0},{\bf{k}})$ is the 
(temperature independent) two-point spectral function given by
\beq
\label{eqn3}
\rho(k_0,\v{k})=2\pi\e(k_0)\de(k_0^2-\w_k^2)
\eeq
where $\e(k_0)$ is the sign function and
\beq
\label{eqn4}
\w_k^2=\v{k}^2+m^2
\eeq

However, the propagators need to be regularised because eventually we want to 
take the Fourier transform in time by taking the limits 
$t_{i}\rightarrow -\infty, t_{f}\rightarrow +\infty$ (to get energy 
conservation) and this is ill-defined since the integrands are linear 
combinations of complex exponentials. For this reason we write the $\delta$
distribution in its regularised form
\beq
\label{eqn5}
\de(k_0\mp\w_k)=\frac 1{2i\pi}\[\frac 1{k_0\mp\w_k-i\e}-
\frac 1{k_0\mp\w_k+i\e}\]
\eeq
and thus $\rho(k_{0},\v{k})$ in \eq{eqn3} can be written as
\beq
\label{eqn6}
\rho(k_0,\v{k})=\frac i{2\w_k}\sum_{r,s=\pm1}\frac{rs}{k_0-s\w_k+i\e r}
\eeq
The regularised propagator can be written as 
\beq
\label{eqn7}
D_R^c(t,\v{k})=D_R^>(t,\v{k})\th_c(t)+D_R^<(t,\v{k})\th_c(-t)
\eeq
and obeys the KMS condition
\beq
\label{eqn8}
D_R^>(t-i\b,\v{k})=D_R^<(t,\v{k})
\eeq

Momentum integration in the complex $k_{0}$-plane will give for
 $D_R^>(t,\v{k})$, with $t$ defined in the region $-\beta\leq\mbox{Imt}\leq 0$
\beq
\label{eqn9}
D_R^>(t)=\frac 1{2\w_k}\sum_{\e=\pm}\[\th(\e)+n(\w_k-i\e s\xi)\]
e^{-i\e w_kt-\xi st}-\frac 1{2\w_k\b}\sum_{\eta\ge 1}X_{\eta} e^{-s\w_{\eta} t}
\eeq
where $s={\rm sign}\[Re(t)\]$, $\xi$ is the regulator and $X_\eta$ is the sum
\beq
\label{eqn10}
X_\eta=\sum_{r,s=\pm1}\frac{rs}{i\w_\eta-s\w_k+i\xi r}=\frac{8\xi\w_\eta\w_k}
{[(i\w_\eta-\w_k)^2+{\xi}^2][(i\w_\eta+\w_k)^2+{\xi}^2]}
\eeq
where $\w_\eta=\frac{2\pi\eta}{\b}=2\pi\eta T$ denotes the Matsubara 
frequencies.
This term arises from the residues of the distribution function when we 
integrate $k_0$
in the complex plane. We will discuss its contribution later. 

\subsection{The bubble term}
The rest of the paper will be the calculation of the bubble diagram,
which is nothing else but the product of two propagators. In order to
justify its relevance, we will briefly mention  where it comes from.
We consider the two scalar field theory described by the
Lagrangian 
\beq
\label{eqn11}
{\cal L} [\phi,\eta]=\frac 1{2}\partial_{\mu}\eta\partial^{\mu}\eta-\frac 1{2}m^2{\eta}^2-\frac 1{2}g\phi{\eta}^2+{{\cal L}_0}
\eeq 
where ${{\cal L}_0}$ denotes the free Lagrangian for $\phi$.
If we integrate out the
$\eta$-field fluctuations and use a one-loop approximation we find
that the generating functional can be expressed as 
\beq
\label{eqn12}
{\cal Z}=\int_C{\cal D}\phi e^{iS_0 [\phi ]+iS_{eff} '[\phi ]}
\eeq
 where $S_{eff} '$ is given by
\beq
\label{eqn13}
S_{eff} '[\phi ]=\frac i{2}\mbox{Trln}[1-g\Delta_c (x,x')\phi (x')]
\eeq
and $\Delta_c (x,x')$ is the propagator for the $\eta$ field. 
Expanding the logarithm we get
\beq
\label{eqn14}
S_{eff} '[\phi ]=\sum_{p=1}^{\infty }S_{eff}^{(p)}
\eeq
where $p$ denotes the number of the propagators.
In this expression we will concentrate on the first non-local
term which is the quadratic part  $S_{eff}^{(2)}$ of the expansion and is 
given by
\beq
\label{eqn15}
S_{eff}^{(2)}=\frac{-ig^2Tr(\Delta_c(x,x')\phi(x')\Delta_c(x',x)\phi(x))}{4}
\eeq
One can permute the order of the elements inside the trace using the identity
\begin{eqnarray}
\label{eqn16}
\phi (x)\Delta_c(k) = \Delta_c(k+p)\phi (x) 
\end{eqnarray}
which is equivalent to the Taylor expansion of Fraser \cite{fras} for moving 
momentum operators to the left of functions depending on $x$, when we identify 
$p_{\mu}=-i\partial_{\mu}$.
Thus the quadratic part of $S_{eff} '[\phi ]$ can be rewritten as
\begin{eqnarray}
\label{eqn17}
S_{eff}^{(2)}=-\frac{1}{4}\int d^{4}x\int d^{4}x'\phi (x)iB(p,\beta)\phi
(x')
\end{eqnarray}
 with $iB(p,\beta )$ being the bubble 
term given in terms of the propagators as
\beq
\label{eqn18}
iB(p,\beta )=g^2\int\frac{dk_0}{2\pi}
\int\frac{d^3\v{k}}{{(2\pi)}^3}\Delta_c (k,m)\Delta_c (k+p,m)
\eeq
Separating the time dependence which interests us at finite temperature, 
$S_{eff}^{(2)}$ is written
\begin{eqnarray}
\label{eqn19}
S_{eff}^{(2)}&=&-\frac{ig^2}{4}\int_{t_i}^{t_i-i\beta}dt\int\frac{d^{3}\v{p}}
{{(2\pi)}^3}
\phi (\v{p},t)\nonumber\\
&\times&\int\frac{d^3\v{k}}{{(2\pi)}^3}\int_{t_i}^{t_i-i\beta}dt'
\Delta_c (\v{k};t,t')
\Delta_c(\v{k}+\v{p};t',t)\phi (\v{p},t')
\end{eqnarray}
and the fields $\phi$ are periodic over the time path $[t_i, t_i-i\beta]$. 
This derivative expansion of the bubble
term is well established at zero temperature. At finite temperature
this is not so, since we will be expanding our theory around an ill-defined
 point. This can be seen for example if we look at the ${\it{\Delta_{11}}}$
component of the propagator of our theory which is given by 
\beq
\label{eqn20}
\Delta(k,m)=\frac{1}{k^2-m^2+i\e}-2i\pi n(k_0)\delta(k^2-m^2)
\eeq
Substituting the propagator in \eq{eqn17}, the temperature-dependent real 
part of the quadratic thermal
effective action is given by the following expression which is nonanalytic 
at the zero four-momentum limit \cite{das}
\begin{eqnarray}
\label{eqn21}
\mbox{Re}({S_{eff}}^{(2)}[\phi ])=-\frac{g^2}{32\pi^2}\int d^4x \left\{\int_{0}^{\infty}dkf(k)\phi(x)\right\}\phi(x)
\end{eqnarray}
with
\begin{eqnarray}
\label{eqn22}
f(k)=\frac{kn(\w)\mbox{Re(ln}R)}{\w(-\nnabla^2)^{1/2}}
\end{eqnarray}
where $\w^2 =\v{k}^2+m^2$ and
\beq
\label{eqn23}
R=\prod_{r,s=+1,-1}(\partial_0^2 -\nnabla^2+2is\w\partial_0+2irk(-{\nnabla}^2)^{1/2})^r
\eeq
This result suggests that the derivative expansion breaks down at finite
temperature, due to the problematic product of the two
delta functions contained in the bubble. If the derivative expansion 
of the effective action is
not rigorously possible, then the definition of an effective
potential, the lowest order term in such an expansion, is not unique.
This would have consequences in any kind of study
concerning symmetry breaking and restoration, unless there is a formal
way to overcome these pathologies and have a well defined derivative expansion.

\subsection{Effective potential term}

We use Le Bellac and Mabilat's formulation \cite{belmab} to prove explicitly 
the $t_{i}$-independence 
of the effective potential in the case of two propagators and one external 
time $t_{0}$ (first term of the bubble diagram) 
when adding both contributions from the 
horizontal and vertical path. Since we are interested in time dependent 
fields, we will concentrate on the time integral of the bubble term 
in \eq{eqn19}. This is given by
\beq
\label{eqn24}
G_C^R=\int_C D^c_R(t_0-t_1)D_R^c(t_0-t_1)dt_1
\eeq 
where C will be in the region $[t_{i},t_{0}]$ and $[t_{0},t_{i}]$ for the
horizontal path and $[t_{i},t_{i}-i \beta]$ for the vertical one as shown 
in Fig.1. Using the representation for the propagators of \eq{eqn9}, the 
integrations over the two paths give $G_H^R$ for the horizontal and $G_V^R$ for
the vertical one \cite{belmab}
\beq
\label{eqn25}
G_H^R=-\frac{A_H}{(2\w)^2}
(\frac{1-e^{-\b\w(\e_1+\e_2)}e^{2i\xi \b}}{i\w(\e_1+\e_2)+2\xi })
\[1-e^{-i\w(\e_1+\e_2) (t_0-t_i)}e^{2\xi (t_i-t_0)}\]
\eeq
\beq
\label{eqn26}
G_V^R=-\frac{A_V}{(2\w)^2}
(\frac{e^{\b\w(\e_1+\e_2)}e^{-2i\xi \b}-1}{i\w(\e_1+\e_2)+2\xi })
\[e^{-i\w(\e_1+\e_2) (t_0-t_i)}e^{2\xi(t_i-t_0)}\]
\eeq
where 
\beq
\label{eqn27}
A_{H,V}=\sum_{\e_1,\e_2}\[\th(\pm\e_1)+n(\w-i
\xi \e_1)\]
\[\th(\pm\e_2)+n(\w-i\xi \e_2)\]
\eeq
are $t_{i}$-independent coefficients.
The KMS relation which reads as \beq
\label{eqn28}
\[\th(\e)+n(\w-i\xi \e)\]=e^{\e\b\w}e^{-i\xi \b}\[\th(-\e)+n(\w-i\xi \e)\]
\eeq
gives for the coefficients $A_{H}$ and $A_{V}$
\beq
\label{eqn29}
A_H=e^{(\e_1+\e_2)\b\w}e^{-2i\xi \b}A_V
\eeq
 Adding 
the $t_{i}$-dependent terms of both paths \eq{eqn25} and \eq{eqn26}
and using the KMS 
condition of \eq{eqn28}, we find that they cancel. Therefore,
 it is essential that we
add the vertical contribution to ensure the $t_{i}$-independence of our
result. The remaining part of the sum is given by
 \beq
\label{eqn30}
G^R=G^R_H+G^R_V=-\frac 1{(2\w)^2}\frac{(A_H-A_V)}{i\w(\e_1+\e_2)+2\xi }
\eeq
and using the definitions of $A_H$ and $A_V$ from \eq{eqn27} and identities 
of the $\theta$ functions, the sum is written as
 \beq
\label{eqn31}
G^R_H+G^R_V=-\frac 1{(2\w)^2}
\left\{\frac{(1+2n(\w))i\w+2i\xi ^2n'(\w)}{-\w^2-\xi ^2}\right\}
+\frac{i\b n(\w)(1+n(\w))}{2\w^2}
\eeq
In \eq{eqn31} the first term is the result of \eq{eqn30} for $\e_1=\e_2$ and 
the second term is the result for $\e_1+\e_2=0$. 
Taking the limit of the regulator $\xi$ to zero at the end, we have
\beq
\label{eqn32}
G^R_H+G^R_V=\frac{i}{4\w^3}(1+2n(\w))+\frac{i\b n(\w)(1+n(\w))}{2\w^2}
\eeq
The previous result agrees completely with the ITF result for the effective
potential \cite{eva3}, which proves the consistency of our theory to 
this order.
Now we examine the cases of taking different limits for the $t_{i}$
and the regulators and try to explain their physical meaning.
\be
\item We take the limit $t_{i}\rightarrow -\infty$ (keeping the
regulators finite) which should give us the real time formalism. 
Since the total sum is $t_{i}$-independent, the limit of 
$t_{i}\rightarrow -\infty$ is already given by \eq{eqn31} in which the 
regulator can be taken to zero since there is no need for it any more 
after the 
limit has been performed. We notice from \eq{eqn26} that keeping 
the regulators finite
the vertical part vanishes in this limit, recovering thus the RTF and the
total sum is being given by the $t_{i}$-independent contribution of the
horizontal part. In the case of ${\it unregularised}$ propagators,
the vertical contribution contains a $t_{i}$-independent term of the form
$$
\frac\b i\frac{1}{2\w}n(\w)(1+n(\w))2\pi \de (k^2-m^2)
$$
which in the regularised approach is hidden in the two horizontal parts 
of the contour, as seen in the last part of \eq{eqn31}. We see that, 
to this order, the regularised 
formalism
is dealing with the pathologies of the problematic delta functions, recovering
 RTF in the appropriate limit.
\item Now keeping $t_{i}$ finite, we take the zero limits of the regulators
in different orders and find
$$
G^R_H(\e_1+\e_2=0,\xi\rightarrow 0)=
G^R_H(\xi\rightarrow 0,\e_1+\e_2=0)=0
$$
and the only contribution comes from the vertical part in the limit
$\e_1+\e_2=0, \xi\rightarrow 0$ which is the second term of \eq{eqn31}
$$
\lim_{\xi \rightarrow0,\e_1+\e_2=0} G=\frac{i\b n(\w)(1+n(\w))}{2\w^2}
$$
This limit is one part of the full effective potential of \eq{eqn32} 
as expected since it only corresponds to the limit of equal and opposite 
energies 
$\w$ for the two propagators $(\e_1+\e_2=0)$. 
\ee

\subsection{The second derivative term}

Now the same formalism will be used for the derivation of the {\it second 
derivative} term of the effective action in our 1-loop case, where
only time-dependent fields are considered. Because now the sum
will also contain terms polynomial in $(t_{i}-t_{0})$ as well as 
exponential ones, the equivalence with the RTF
and ITF is less straightforward. We will show that the RTF limit can be
extracted in this case without problems of divergences as long as we keep
the regulators finite and take them to zero after the limit has been done.
The second derivative term will look like 
\beq
\label{eqn33}
\Gamma^{(2)}_2=\int_C dt_1 D^c_R(t_0,t_1)(t_0-t_1)^2D^c_R(t_1,t_0)
\eeq
where we have omitted the $\frac 1{2}\partial_t^2\phi(t_0)$ factor of this 
term in the expansion. 
Using the definition of \eq{eqn1}, $\Gamma^{(2)}_2$ can be written as
 \begin{eqnarray}
\label{eqn34}
\Gamma^{(2)}_2=\int_C dt_1 \int\int\frac{dk_0 dk_1}{(2\pi)^2}&
(\th_c(t_0,t_1)+n(k_0))(\th_c(t_1,t_0)+n(k_1))\nonumber\\
 &\times(t_1-t_0)^2 e^{-i(k_0-k_1-i\e)(t_0-t_1)}\rho(k_0)\rho(k_1)
\end{eqnarray} 
where the regulator $\e$ is used so that the limit of
$t_{i}\rightarrow -\infty$ can be taken without problems. In the end it will
be set to zero. The other two regulators $\e_1$ and $\e_0$ in the
delta
functions of $\rho(k_0)\rho(k_1)$ make sure that no problems appear in the 
equal energy (mass) case $k_0=k_1=w$.\\
Now we first perform the $dt_1$ integration and then ``absorb'' 
the $(t_1-t_0)^{2}$ 
term by differentiating the result with respect to $k_0$. If we name the 
time integrals over the two paths  as $I_C$, we then write
\beq
\label{eqn35}
\int_Cdt_1(t_1-t_0)^2\th_c(t_1,t_0)e^{-i(k_0-k_1-i\e)(t_0-t_1)}=
i^2\frac{\partial^2}{\partial k_0^2}I_C
\eeq
Substituting this formula into our general expression \eq{eqn34}, the 
contributions from the different paths can now be written 
\beq
\label{eqn36}
\Gamma_H^{(2)}=\int\frac{dk_0 dk_1}{(2\pi)^2}(n(k_0)-n(k_1))\rho(k_0)\rho(k_1)
i^2\frac{\partial^2}{\partial k_0^2}I_H
\eeq
\beq
\label{eqn37}
\Gamma_V^{(2)}=\int\frac{dk_0dk_1}{(2\pi)^2}
\[n(k_0)(n(k_1)+1)]\rho(k_0)\rho(k_1)
i^2\frac{\partial^2}{\partial k_0^2}I_V
\eeq
In $\Gamma_H^{(2)}$ the $\th^2$ term vanishes due to the opposite sign of 
its time arguments. The $n^2$ term vanishes due to the cancellation between 
the two horizontal paths. In $\Gamma_V^{(2)}$ one of the $\th$-functions 
always vanishes due to the choice of $t_0$ on the horizontal path.

The time integrations $I_H,I_V$ over the two paths give
$$
I_H=i\frac{(1-e^{-i(k_0-k_1-i\e)(t_0-t_i)})}{k_0-k_1-i\e}
$$
$$
I_V=ie^{-i(k_0-k_1-i\e)(t_0-t_i)}\times\frac{(1-e^{\b(k_0-k_1-i\e)})}{k_0-k_1-i\e}
$$

The analytical calculation of the different path contributions is quite 
complicated since it involves first and second order residues and therefore 
derivatives of the distribution functions. We performed the momentum 
integrations and then took the same limits of our variable $\Delta t=t_i-t_0$ 
and 
of the regulators as before to check the consistency of our method for the 
second derivative term.
\be
\item We took the $\Delta t\rightarrow -\infty$ limit keeping the regulators 
finite.
In the total sum the $\Delta t$-dependence appears in terms 
like $\Delta t^ne^{\e\Delta t}$ and 
$\Delta t^n$ $(n=0,1,2)$. These terms could cause divergences 
in the $\Delta t\rightarrow -\infty$ limit but they disappear once we 
include the vertical part in our calculation. Our result is 
{\it independent} of the order in which the regulators are taken to zero in 
the end and is given by a finite term coming from the horizontal part
$$
\lim_{t_i\rightarrow -\infty}\Gamma^{(2)}=-\frac i8\frac{(1+2n(\w))}{\w^5}
$$
This term looks like the first order term of the effective potential divided 
by $\w^2$, as it can be seen from \eq{eqn32}, which is sensible since it is 
essentially the first correction due to the second derivative.
\item Our second limit is $t_i=t_0, \e=0$ in order to try to recover the ITF 
result
($t_i=t_0=$ finite and $\e$ is not needed any more since $t_i$ is finite). 
This proved to be also independent of the order of the zero limits of the 
regulators. We obtained 
\beqa
\lefteqn{\Gamma^{(2)}(t_i=t_0, \e=0;\e_1=0;\e_0=0)=}\\
& & -\frac i{8\w^5}\[(2n(\w)+1)-\beta\w(2n(\w)(n(\w)+1)+1)\\
& & \mbox{}+\b^2\w^2(2n(\w)+1)+\frac{4\b^3\w^3n(\w)(n(\w)+1)}{3}]
\eeqa
which is consistent with the derivation of the second derivative term
 in the ITF formalism \cite{eva3}.
\ee

\section{Alternative method}
Another possible way of performing our calculation is to consider the full 
Taylor series of the field but do the actual expansion and study the 
individual terms in the end.
We will generalise our method considering different energies 
$(\w$ and $\Omega)$ in the delta functions of \eq{eqn3} for each propagator 
of the 
bubble term. In this way we will check the analyticity limits of the full 
derivative term by taking the limits 
$\Omega\rightarrow \pm\w$ ($\v{\nnabla}\rightarrow0$) 
and $\partial_t\rightarrow0$ in different orders at the end of the calculation.

The expanded field can be written as 
$$
\phi(t_1)=\sum_{n=0}^\infty\left. \frac 1{n!}(t_1-t_0)^n\frac{\partial^n}{\partial_t^n}\phi(t)\right\vert_{t=t_0}\!\ =\left. e^{(t_1-t_0)\partial_t}\phi(t)\right\vert_{t=t_0}
$$
The full derivative term of the field inserted between the two propagators, 
which is the last time integral in \eq{eqn19}, looks like
\beq
\label{eqn38}
\Gamma^{(B)}=\int_C dt_1 D^c_R(t_0,t_1)\left. e^{(t_1-t_0)\partial_t}\phi(t)\right\vert_{t=t_0}  D^c_R(t_1,t_0)
\eeq
This term acts as an energy-shift by $-i\partial_t$ in the exponentials 
of the propagators making the time-integrals over the paths $I_C$ 
of \eq{eqn35} look like 
\beq
\label{eqn39}
\int_Cdt_1\th_c(t_1,t_0)e^{-i(k_0-k_1-i(\e+\partial_t))(t_0-t_1)}= I_C '
\eeq
Performing the time integration for the horizontal and vertical path as 
before, we get
$$
I_H '=i\frac{(1-e^{-i(k_0-k_1-i(\e+\partial_t))(t_0-t_i)})}{k_0-k_1-i(\e+\partial_t)}
$$and
$$
I_V '=ie^{-i(k_0-k_1-i(\e+\partial_t))(t_0-t_i)}\times\frac{(1-e^{\b(k_0-k_1-i(\e+\partial_t))})}{k_0-k_1-i(\e+\partial_t)}
$$
The energy integration gives us the full bubble term as a sum over the 
two paths written in terms of $\Delta t=t_i-t_0$
$$
\Gamma^{(B)} =\Gamma_H ^{(B)} + \Gamma_V ^{(B)}
$$ with  
\beq
\label{eqn40}
\Gamma_H ^{(B)} =\sum_{\pm\w,\Omega}\frac{in(\w-i\e_0)n(\Omega-i\e_1)}{4\w\Omega}\times[\frac{(e^{\beta(\w+\Omega-i(\e_0+\e_1))}-1)(1-e^{-iA\Delta t})}{A}]
\eeq
and 
\beq
\label{eqn41}
\Gamma_V ^{(B)} =\sum_{\pm\w,\Omega}\frac{in(w-i\e_0)n(\Omega-i\e_1)}{4\w\Omega}\times[\frac{e^{-iA\Delta t}e^{-i\beta(\e +\partial_{t})}(e^{\beta A }-1)}{A}]
\eeq
where $$A=\w+\Omega -i(\e_1+\e_0-\e-\partial_t)$$ 
Now we can check the analyticity of our result keeping $\Delta t$ finite. 
We expand the distribution functions and take the limits of our regulators 
to zero (we can do that since we keep $\Delta t$ finite). 
If we take the limits $\Omega\rightarrow \pm\w$ and $\partial_t\rightarrow 0$,
 we get finite and independent of the order of the limits results. 
The full derivative expansion of the bubble term, therefore, 
is analytical in this limit and is
\beq
\label{eqn42}
\Gamma_H ^{(B)}(\Omega\rightarrow \pm\w, \partial_t\rightarrow0) =i
 \frac{(2n(\w)+1)(1-cos(2\w\Delta t))}{2\w^3}
\eeq
for the horizontal case and
\beq
\label{eqn43}
\Gamma_V ^{(B)}(\Omega\rightarrow \pm\w, \partial_t\rightarrow0) =i
\frac{(2n(\w)+1)cos(2\w\Delta t)}{2\w^3}+i
\frac{\beta n(\w)(n(\w)+1)}{w^2}
\eeq
for the vertical one.
\be
\item Now we consider the limit $t_i=t_0$ in \eq{eqn42} and \eq{eqn43}. 
This gives 
\beq
\label{eqn44}
\Gamma_H ^{(B)}=0
\eeq
and
\beq
\label{eqn45}
\Gamma_V ^{(B)}=i\frac{(2n(\w)+1)}{2\w^3}+i\frac{\beta n(\w)(1+n(\w))}{w^2}
\eeq
This is exactly the result for the effective potential using the ITF 
formalism, as expected since it is the zeroth time and space-derivative term, 
when $t_i=t_0$. It also agrees with our previous result in \eq{eqn32} 
of the effective potential after we set the regulators to zero. (In \eq{eqn32} 
the result differs by a factor of $1/2$ due to the fact that we have 
initially considered same energies $\w$ for the propagators and this 
corresponds to half of the result of the effective potential of \eq{eqn45} 
when different energies are assumed).
\item Now we take the limit $\Delta t\rightarrow -\infty$ and we will 
check the analyticity again. In this case only the first 
$\Delta t$-independent part of $\Gamma_H ^{(B)}$ survives and 
taking the regulators to zero after the limit has been performed, we get 
\beq
\label{eqn46}
\Gamma ^{(B)}(\Delta t\rightarrow -\infty)= \sum_{\pm\w,\Omega}\frac i{4\w\Omega}\frac{(n(\w)-n(-\Omega))}{(\w+\Omega+i\partial_t)}
\eeq
The analyticity check for \eq{eqn46} gives finite but different results 
for different orders of performing the limits 
$(\Omega\rightarrow \pm\w, \partial_t\rightarrow0)$. We found that 
performing the time-derivative $(\partial_t\rightarrow0)$ limit first 
and the spatial-derivative $(\Omega\rightarrow \pm\w)$ afterwards, we 
had the usual effective potential term of \eq{eqn45}
$$
\lim_{\partial_t\rightarrow 0,\Omega\rightarrow \pm\w}\Gamma ^{(B)}=i\frac{(2n(\w)+1)}{2\w^3}+i\frac{\beta n(\w)(1+n(\w))}{w^2}
$$
but reversing the order of the limits gave us only the first term of our 
previous result
$$
\lim_{\Omega\rightarrow \pm\w,\partial_t\rightarrow 0}\Gamma ^{(B)}=i\frac{(2n(\w)+1)}{2\w^3}
$$
We see that although we don't have divergence problems in taking the limits 
in both orders, approaching the zero from the space-derivative first seems 
to produce only part of the full result in agreement with the result of Evans 
using ITF \cite{eva3}. Now we take only the spatial 
derivative to zero in the $\Delta t\rightarrow -\infty$ case of \eq{eqn46} 
which gives
$$
\lim_{\Omega\rightarrow \pm\w}\Gamma ^{(B)}=2i\frac{(2n(\w)+1)}{\w(4\w^2+\partial_t ^2)}
$$
If we now expand our result in powers of the time-derivative $\partial_t$, 
we get the zeroth order term of our analyticity check and a second order 
term of the form
$$
-\frac{i(2n(\w)+1)}{8\w^5}(\partial_t ^2)
$$
This is exactly  the second order time-derivative term derived in this limit 
using our previous method in section 2.4. We have to note that in 
the case of $\Delta t\rightarrow -\infty$, there is no term linear 
in $\partial_t$. 
\ee 

Now we perform the same expansion in powers of the time-derivative up to 
the second order but for a general finite $\Delta t$, for both horizontal 
and vertical paths in \eq{eqn40} and \eq{eqn41} and take the limits 
$(\Omega\rightarrow \pm\w)$ to get the effective potential and the higher 
derivative terms. We identify the terms as follows
\be
\item $\partial_t^0$ term

\beq
\label{eqn47}
\Gamma_H ^0=-i\frac{(2n(\w)+1)}{4\w^3}(e^{2i\w \Delta t}+e^{-2i\w \Delta t}-2)
\eeq

\beq
\label{eqn48}
\Gamma_V ^0=i\frac{(2n(\w)+1)}{4\w^3}(e^{2i\w \Delta t}+e^{-2i\w \Delta t})+i\frac{\beta n(\w)(n(\w)+1)}{\w^2}
\eeq

\item $\partial_t^1$ term

\begin{eqnarray}
\label{eqn49}
\Gamma_H ^1=\frac{(2n(\w)+1)}{8\w^4}[(e^{2i\w \Delta t}-e^{-2i\w \Delta t})-2i\w(\Delta t)(e^{2i\w \Delta t}+e^{-2i\w \Delta t})]
\end{eqnarray}

\begin{eqnarray}
\label{eqn50}
\Gamma_V ^1&=&-\frac 1{8\w^4}[(2n(\w)+1)(e^{2i\w \Delta t}-e^{-2i\w \Delta t})
\nonumber\\
&-&2\beta\w({(n(\w)+1)}^2 e^{2i\w \Delta t}-n^2 (\w) e^{-2i\w \Delta t})
-4\beta^2 \w^2 n(\w)(n(\w)+1)]\nonumber\\
&+&\frac i{4\w^3}[(2n(\w)+1)(e^{2i\w \Delta t}+e^{-2i\w \Delta t})\nonumber\\
&+&4\beta\w n(\w)(n(\w)+1)](\Delta t)
\end{eqnarray}

We notice the existence of a non-zero $\partial_t$-dependent term unlike the 
zero temperature case where such a term vanishes. This could be related to the
 loss of Lorentz 
invariance in the finite temperature case and could be interpreted as an 
energy shift. The existence of such a linear term might be of great 
physical importance in the study of time-dependent systems.
 Such a term did not exist in the expansion for the $\Delta t$ infinite case,
 where only zero and second order terms in the time-derivative survived. 
This makes sense since in the infinite time limit any interaction with the 
heat bath which gives rise to such linear terms will have been damped.
 Mathematically this term could arise due to the shape of the time contour,
 which in the finite
$\Delta t$ case is non-symmetric. However this is not the case for the 
zero-temperature situation or the non zero temperature one in the infinite 
$\Delta t$ limit where the symmetry of the contour will make any time 
integration of odd terms in the derivative expansion to vanish.  
In the $\Delta t=0$ case this term is equal to 
$$\Gamma ^1=\frac{-i\beta\Gamma ^0}{2}$$ where ${\it{\Gamma ^0}}$ is the 
effective 
potential term given by \eq{eqn47} and \eq{eqn48} in the $\Delta t=0$ limit.

\item $\partial_t^2$ term

\begin{eqnarray}
\Gamma_H ^2 &=&\frac{i(2n(\w)+1)}{16\w^5}[(e^{2i\w \Delta t}+
e^{-2i\w \Delta t}-2)-2i\w(\Delta t)(e^{2i\w \Delta t}-e^{-2i\w \Delta t})
\nonumber\\
&-&2\w^2{(\Delta t)}^2(e^{2i\w \Delta t}+\e^{-2i\w \Delta t})]
\label{eqn51}\\
\nonumber\\
 \label{eqn52}
\Gamma_V^2 &=&\Gamma_{V1}^2+\Gamma_{V2}^2
\end{eqnarray}
with
\begin{eqnarray}
\label{eqn53}
\Gamma_{V1}^2 &=&-\frac i{16\w^5}[(2n(\w)+1)(e^{2i\w \Delta t}+e^{-2i\w \Delta t})
-2\beta\w({(n(\w)+1)}^2 e^{2i\w \Delta t}\nonumber\\
&+&n^2 (\w)e^{-2i\w \Delta t})\nonumber\\
&+&2\beta^2 \w^2({(n(\w)+1)}^2 e^{2i\w \Delta t}-n^2 (\w)e^{-2i\w \Delta t})
\nonumber\\&+&\frac 8{3}\beta^3 \w^3 n(\w)(n(\w)+1)]
\end{eqnarray}
and
\begin{eqnarray}
\label{eqn54}
\Gamma_{V2}^2&=&-\frac 1{8\w^4}[(2n(\w)+1)(e^{2i\w \Delta t}-e^{-2i\w \Delta t})
-2\beta\w({(n(\w)+1)}^2 e^{2i\w \Delta t}\nonumber\\
&-&n^2 (\w)e^{-2i\w \Delta t})-4\beta^2 \w^2 n(\w)(n(\w)+1)](\Delta t)
\nonumber\\
&+&\frac i{8\w^3}[(2n(\w)+1)(e^{2i\w \Delta t}+e^{-2i\w \Delta t})\nonumber\\
&+&4\beta\w n(\w)(n(\w)+1)]{(\Delta t)}^2
\end{eqnarray}

If we take the $t_i=t_0$ limit of our second derivative term, we recover our 
previous derivation of the same term in section 2.4.

\ee

In our calculation we have omitted the contribution of the $X_\eta$ term 
of \eq{eqn10}. This term which arises from the residue of the distribution 
function vanishes since it is proportional to the regulator. 
In the finite $\Delta t$ case the regulators are set
to zero before any limit is taken while in the infinite
$\Delta t$ case they are set to zero once the infinity limit has been 
performed. In both cases this term does not contribute.

\section{The initial time dependence}

In this section we will treat the initial time-dependence of our
problem in a rather more formal way.

Based on Le Bellac and Mabilat's proof of the $t_{i}$-independence of a
regularised Green function \cite{belmab}, we will prove the same for our 
effective potential term. 
Our $t_i$-dependent integrals in this case are
\beq
\label{eqn55}
\int_{t_i}^{t_0} dt_1G_R(t_1,t_0)G_R(t_1,t_0)+\int_{t_0}^{t_i-i\beta} dt_1G_R(t_1,t_0)G_R(t_1,t_0)
\eeq
Differentiating the first term with respect to $t_i$ we get
$$-G_R^<(t_i,t_0) G_R^<(t_i,t_0)$$
Repeating for the second term we now get
$$G_R^>(t_i-i\beta,t_0) G_R^>(t_i-i\beta,t_0)$$
Using the KMS condition for thermal equilibrium
$$G_R^>(t-i\beta)=G_R^<(t)$$
we see that these terms cancel.
If we repeat the same method for the higher derivative terms of the 
field of the bubble case, we get a non-zero result which shows explicitly 
the $t_i$-dependence of these terms. The same analysis for the second 
derivative term gives
\begin{eqnarray}
\label{eqn56}
G_R^<(t_i,t_0)[-\beta^2-2i\beta(t_i-t_0)]G_R^<(t_i,t_0)
\end{eqnarray}
If we generalise in the case of the m-th derivative term, 
the $t_i$ dependence of the derivative with respect 
to $t_i$ will have the form
\begin{eqnarray}
\label{eqn57}
G_R^<(t_i,t_0)[\sum_{k=1}^m\prod_{j=0}^{k-1}\frac{(m-j)}{k!}(-i\beta)^k(t_i-t_0)^{m-k}]G_R^<(t_i,t_0)
\end{eqnarray}
We see that the individual terms of the expanded field are clearly 
$t_i$-dependent even in the case of $t_i=t_0$, where the highest 
order $\beta$-term survives in the previous sum. 
This is somehow expected since a truncated expansion of the field makes 
it no longer periodic. On the other hand if we have a periodic field 
$\phi(t)$ in equilibrium, the derivative with respect to $t_i$  discussed 
earlier will give
\beq
\label{eqn58}
-G_R^<(t_i,t_0)\phi(t_i)G_R^<(t_i,t_0)+G_R^>(t_i-i\beta,t_0)\phi(t_i-i\beta)G_R^>(t_i-i\beta,t_0)
\eeq
which is zero for periodic fields and regularised propagators 
obeying the KMS condition.

\section{Conclusions and possible applications}
We found that using this closed time path formalism we can avoid the 
pathologies in the RTF and derive the ITF limit as well, both in the
case of the effective potential and in the second derivative correction of the 
bubble diagram. The fact that we can cancel the divergence in our
effective action using our prescription, hence providing us with a
formalism which allows us to compute quantum corrections to the
effective potential is the key point of our paper. However we found that the 
inclusion of the vertical path and the careful treatment of the regulators
are essential for the cancellations to happen. 

We showed a general way to 
compute higher derivative terms in the bubble and 
derived the complete bubble term. We checked its analyticity for finite and 
infinite time differences $\Delta t$ and found different limits in the second
case. The non zero, linearly dependent on the time derivative of the 
field term found in the finite $\Delta t$ case, may be related to the
loss of translational invariance at finite temperature. The physical meaning of
such a term and in particular its sign and whether it is complex or real 
may be important in the study of time-dependent systems. Gribosky and 
Holstein \cite{grib} do not find such a linear term in their 
expansion of derivatives of the field. Motivated by the use of Feynman 
parametrisation at zero temperature \cite{weld3} 
they calculate the vacuum polarisation 
diagram using ITF but extending to continuous $p_0$ first and evaluating 
the mode sum afterwards. They compare their result with Dittrich's \cite{dit} 
background field method of calculating the effective Langrangian at 
finite temperature and in both cases 
there is no linear term unlike our case which appears for any finite 
$\Delta t$. 

The extension of our calculation to higher derivative terms and to 
space-dependent fields will give a full effective action whose importance in
field theory was discussed earlier. Our calculation
may be performed for higher order diagrams in the expansion of the
one-loop effective action, but this is beyond the scope of this paper.
We have considered a two real scalar field theory, but we
could in principle use our method in different models, such as a Yukawa or a 
gauge theory or even consider systems with time-dependent parameters. The 
possibility of evaluating quantum corrections can be 
more directly applied to phase transitions, where they
may indicate us something about the order of the transition.

\section{Acknowledgements}

We would like to thank T. S. Evans and R. J. Rivers for many helpful 
discussions, A. Gomez Nicola for his interesting remarks and B. D. Wandelt 
for his help with Mathematica. 
This work has also been supported in part by the European 
Commission under the Human Capital and Mobility programme, contract 
number CHRX-CT94-0423.

\pagebreak


\begin{thebibliography}{99}

\bibitem{belmab}M. Le Bellac and H. Mabilat, Z. Phys. C{\bf 75}, 137, 1997\\
H. Mabilat, Z. Phys. C{\bf 75}, 155, 1997
\bibitem{bel}M. Le Bellac, ``Thermal Field Theory'', Cambridge University 
Press, 1996
\bibitem{Keld}L. V. Keldysh, Sov. Phys. JETP {\bf 20}, 1018, 1964\\
J. Schwinger, J. Math. Phys. {\bf 2}, 407, 1961\\R. A. Craig, Ann. Phys. 
{\bf 40}, 416, 1966
\bibitem{Matsu}T.Matsubara, Prog. Theor. Phys. {\bf 14}, 351, 1955
\bibitem{ander}G. Anderson and L. Hall, Phys. Rev, D {\bf 45}, 2685, 1992\\M. 
E. Carrington, Phys. Rev, D {\bf 45}, 2993, 1992 
\bibitem{kaj}K. Kajantie and J. Kapusta, Ann. Phys. C{\bf 160}, 477, 1985
\bibitem{arno}P. Arnold and L. M. Lerran, Phys. Rev, D {\bf 36}, 581,  1987
\bibitem{cole2}S. Coleman, ``Aspects of Symmetry'', Cambridge University 
Press, 1985\\R. J. Rivers, ``Path 
integral methods in quantum field theory'', Cambridge University Press, 1987
\bibitem{jac}R. Jackiw, Phys. Rev, D {\bf 9}, 3320\\S. Weinberg, Phys. Rev, 
D {\bf 9}, 3357, 1974
\bibitem{cole1}S. Coleman, Phys. Rev, D {\bf 15}, 2929, 1977\\C.Callan and S.Coleman, Phys. Rev, D {\bf 16}, 1762, 1977
\bibitem{cole3}S. Coleman and E. Weinberg, Phys. Rev, D {\bf 7}, 1883, 1973 
\bibitem{das}A. Das and M. Hott, Phys. Rev, D {\bf 50}, 6655, 1994 
\bibitem{grib}P. S. Gribosky and B. R. Holstein, Z. Phys. C{\bf 47}, 205, 1990
\bibitem{sili}V. P. Silin, Sov. Phys. JETP {\bf 11}, 1136, 1960\\
D. J. Gross, R. D. Pisarski and L. G. Jaffe, Rev. Mod. Phys. {\bf 53}, 43, 1981
\bibitem{weld1} H. A. Weldon, Phys. Rev, D {\bf 28}, 2007, 1983
\bibitem{evans}T. S. Evans, Can. J. Phys. {\bf 71}, 241, 1993
\bibitem{weld2}H. A. Weldon, Phys. Rev, D {\bf 26}, 2789, 1982
\bibitem{eva1}T. S. Evans, Z. Phys. C{\bf 36}, 153, 1987\\T. S. Evans, 
Z. Phys. C{\bf 41}, 333, 1988
\bibitem{weld3}H. A. Weldon, Phys. Rev, D {\bf 47}, 594, 1993 
\bibitem{mill}R. Mills, ``Propagators for Many-Particle Systems'', Gordon 
and Breach, 1969
\bibitem{fras}C. M. Fraser, Z. Phys. C{\bf 28}, 101, 1985
\bibitem{eva3}T. S. Evans, ``Derivative expansions of Euclidean thermal 
effective actions'', (in preparation)
\bibitem{dit}W. Dittrich, Phys. Rev, D {\bf 19}, 2385, 1979

\end{thebibliography}
\end{document}